\documentstyle[aps,manuscript]{revtex}   

\begin{document}

\draft
\title{A semiclassical theory of dissipative Henon-Heiles system}
\author{Bidhan Chandra Bag and Deb Shankar Ray}
\address{Indian Association for the Cultivation of Science,
Jadavpur, Calcutta 700 032, INDIA.}
\maketitle
\begin{abstract}
A semiclassical theory of dissipative Henon-Heiles system is proposed. Based on
$\hbar$-scaling of an equation for evolution of Wigner quasiprobability
distribution function in presence of dissipation and thermal diffusion, we
derive a semiclassical equation for quantum fluctuations, governed by the dissipation
and the curvature of the classical potential. We show how the initial quantum
noise gets amplified by classical chaotic 
diffusion which is expressible in terms
of correlation of stochastic fluctuations of the curvature of the potential
due to classical chaos and ultimately settles down to equilibrium 
under the influence of dissipation. We also establish that there exists a 
critical limit to the expansion of phase space. The limit is set by chaotic
diffusion and dissipation. Our semiclassical analysis is corroborated
by numerical simulation of quantum operator master equation.
\end{abstract} 

\vspace{2.5cm}

\pacs{{\bf Keywords}: Dissipative quantum system, semiclassical approximation, 
classical chaos, Henon-Heiles Hamiltonian.}

\newpage

\section{Introduction}

\vspace{0.5cm}

The influence of dissipation on quantum dynamics of classically chaotic systems
has been one of the key issues in nonlinear dynamics today. The dynamical
way of dealing with dissipation is to consider a system-heat bath model which
has been the cornerstone for understanding dissipative processes$^{(1)}$ in a wide
range of physical disciplines$^{(2)}$, such as, condensed matter physics, quantum optics,
chemical dynamics etc. The theoretical development in this regard is welldocumented 
in the literature$^{(2-6)}$. When the system, in question, is classically chaotic, one 
envisages a variety of rich physics$^{(7-20)}$ concerning localization and its 
suppression, quantum measurement problem, irreversibility, relaxation, 
decoherence etc. The inferences drawn from these are sometimes extended to
the question of generic quantum chaos. For example, a dissipative quantum system
exhibting chaos in its classical limit was constructed by coupling the 
quantum kicked rotor to a reservoir to obtain dissipative quantum standard map 
by Dittrich and Graham $^{(7)}$. It was observed that even weak damping is capable of
disrupting dynamical localization which suppresses chaotic motion in the 
conservative standard map and thus restores diffusion in action variable in
the timescale of classical relaxation. The effect of quantum correlation on 
classical chaotic behaviour had been illustrated by Sundaram and Milonni$^{(8)}$
by considering a kicked quantum system coupled to a reservoir. An appropriate
choice of potential results in a logistic map with self-consistently 
generated quantum correlations. It has been observed that at intermediate
range of dissipation an irregular behaviour is induced by quantum correlations
even when the classical limit is regular. 
Based on an analysis of quantum Brownian motion in d-dimensions using the 
unified model for diffusion localization and dissipation, Choen$^{(20)}$ has 
proposed a semiclassical strategy at low temperature using Feynman-Vernon 
propagator scheme. It has been demonstrated that different mechanisms
for dephasing emerge for ergodic and nonergodic motions.
In another issue
Bonilla and Guinea$^{(15)}$ have studied a simple model having quantum and classical
degrees of freedom in presence of dissipation. The emergence of chaos in an
open quantum system has also been considered by Spiller and Ralph$^{(16)}$.

The majority of the studies considered above are based on maps, (such
as, standard map or logistic map) which have been the testing ground for
various theories of chaos. We construct here a dissipative version of a 
two-degree-of-freedom continuous system - the Henon-Heiles model$^{(21-23)}$, to
study the evolution of a quantum system in presence of dissipation and 
thermal diffusion. The Henon-Heiles model captures the essential 
generic features of classical chaos in nonintegrable systems and has been
widely applied in the context of astronomy and chemical dynamics over the
last several decades$^{(21-23)}$. 
Based on suitable $\hbar$-scaling of Wigner equation 
which incorporates the effect of dissipation and thermal diffusion, we 
formulate a semiclassical dynamics which is governed by dissipation and 
curvature of the classical potential. The stability of classical motion is
determined by the nature of curvature of the potential which in the
chaotic regime can be considered to be a stochastic process$^{(24, 25)}$.
An appropriate treatment of this stochastic process in terms of the
theory of multiplicative noise yields a Fokker-Planck equation of motion for 
Wigner-function.
We design the initial conditions in terms
of minimum uncertainty wave packets to maximize the classical-quantum  
correspondence and show how the initial wave packet corresponding to a
chaotic trajectory evolves in time, and how the initial quantum noise 
(inherent in minimum uncertainty of wave packet) associated with it
gets amplified by intrinsic
classical stochasticity at the begining to eventually settle down to equilibrium
under the influence of dissipation. We establish that there exists a critical 
limit to the expansion of phase space. The limit is set by chaotic diffusion
and dissipation. Our semiclassical analysis is 
supplemented by quantum simulation of the operator master equation 
to verify the basic theoretical propositions.

The outlay of the paper is as follow: In Sec. II we introduce the quantum 
operator master
equation and the Wigner function equation for an open system. $\hbar$-scaling
of the Wigner equation results in a semiclassical equation governed by the dissipation
due to the surroundings and the curvature of the potential. This is followed by 
van Kampen's treatment of multiplicative noise $^{(26, 27)}$ to deal with 
stochastic fluctuations of the curvature of the
potential which leads to a Fokker-Planck equation.  In Sec III 
the Fokker-Planck equation is adapted to Henon-Heiles system followed by a 
detailed analysis of the problem. In Sec IV a numerical 
simulation of the operator master equation has been carried out to verify
the theorerical propositions. The approximations and their validity with
a summary of the main results have been discussed in Sec V.

\section{Chaotic evolution of an open system; general aspects}
\vspace{0.25cm}

\subsection{Quantum dynamics}
\vspace{0.25cm}

To study$^{(2)}$
the evolution of a quantum system in presence of weak dissipation 
and thermal diffusion we first consider 
the Hamiltonian of an N-degree-of-freedom system $H_0$.

\begin{equation}
H_0 = \sum_{i=1}^N \frac{p_i^2}{2 m_i} +V(\{q_i\}) \; \;,
\; \; i = 1 \cdots N  
\end{equation}

\noindent
where $\{q_i, p_i\}$ represents the coordinates and 
momenta of the N-degree-of-freedom system. $V(\{q_i\})$ is a nonlinear potential such
that the classical version of $H_0$ admits of chaos.

The bare system is then coupled to an environment modeled by a reservoir of 
harmonic oscillators, governed by the following total Hamiltonian 

\begin{equation}
H = H_0 + \hbar \sum_{j}^\infty \Omega_j {b_j }^\dagger b_j +\hbar \sum_{j}^\infty 
\left[k(\Omega_j) b_j +k^\star (\Omega_j) {b_j}^\dagger \right] q \; \;,
\end{equation}

where $b_j$ $({b_j}^\dagger)$ denotes the annihilation (creation) operator for the 
harmonic oscillators which comprise a bath. The third term represents the linear coupling 
of one of the selected degrees of freedom (through co-ordinate $q$) of the system to the bath. 
$k(\Omega_j)$ signifies  
the coupling constant.

It is convenient to invoke first the rotating wave approximation (RWA).
After appropriate elimination of reservoir variables in the usual way using Born and 
Markov approximations we are lead to the following standard reduced 
density matrix 
equation for the evolution of the system$^{(2)}$, 
\begin{equation}
\frac{d \rho}{dt} = - \frac{i}{\hbar} [H_0, \rho] + \frac{\gamma}{2}
(2 a \rho a^\dagger - a^\dagger a \rho - \rho a^\dagger a)
+D(a^\dagger \rho a+ a\rho a^\dagger -a^\dagger a \rho - \rho a a^\dagger) \; \;.
\end{equation}

Here we have expressed the system operators $q,p$ (for $N=1$) in terms of a
harmonic oscillator operators $a$(annihilation) and $a^\dagger$ as 
$q=\frac{1}{\sqrt{2m\omega}} (a+a^\dagger)$ and 
$p=i \sqrt{\frac{m \omega}{2}}(a-a^\dagger)$. Note that the harmonic oscillator
characterized by frequency $\omega$ has nothing to do with the reservoir 
of harmonic oscillators. In the derivation above, one uses a broad
band spectral density function for the reservoir evaluated at $\omega$
to realize the
damping constant $\gamma$ as 
$2 \pi |k(\omega)|^2 g(\omega)$ within a Markovian scheme. $D(=\bar{n} \gamma)$
is the diffusion coefficient and
$\bar{n}(= [exp \left( \frac{\hbar \omega}{k T} \right)-1]^{-1})$
refers to the average thermal photon number of the reservoir. 
The terms analogous to Stark and Lamb shifts are neglected. If more than
one degree of freedom of the system is coupled to the bath 
then the coupling term in Eq.(2) and 
dissipative terms ($\gamma$ and $D$ terms) in Eq.(3) should appropriately
include additional similar contributions [see Eq.(14)].

The first term in Eq.(3) corresponds to the
dynamical motion of the system that
generates Liouville flow and the second term denotes the loss of energy from 
the system to the reservoir, while the last term indicates the diffusion of 
fluctuations of the reservoir into the system of interest. 
The terms containing $\gamma$ 
arise due to the interaction of the system with
the surroundings.

We note that Eq.(3) is a popular form of the operator master equation 
(as derived by 
Louisell$^{(2)}$) which is widely used 
in quantum optics. The equation has also been applied earlier by Dittrich and
Graham$^{(7)}$ in the treatment of dissipative standard map and 
the related problems of chaotic dynamics by others$^{(8,16)}$. 
The Eq.(3) is also general in the sense that we need not ascribe 
any notion of 
regularity or chaoticity in describing the motion governed by the 
Hamiltonian system($H_{0}$). The correlation between different forms of 
operator master equation has been reviewed in Ref$^{(5)}$. All of them, however, 
are not well-suited for numerical simulation. 
Eq.(3) suits this purpose well. We shall
return to this issue in Sec. IV to verify the theoretical propositions.
We note, in passing, that Eq.(3) is based on rotating wave approximation and 
Born-Markov approximation. The latter approximation restricts its validity to 
weak damping limit only.

%\newpage

\subsection{Semiclassical theory}
\vspace{0.25cm}
 
Our next task is to go over from a full quantum operator problem
to an equivalent 
c-number problem described by the Hamiltonian(2). To this end 
we consider the 
quasi-classical distribution function W ($\{q_i\}, \{p_i\}$, t)
of Wigner$^{(28)}$. The time evolution of this phase space function of the dynamical
system characterized by the c-number  variables $\{q_i, p_i\}$ is based on two considerations: First, one takes into account of the usual
dynamical evolution under the influence of potential $V$ as defined in (1). 
The second is the 
dissipative evolution of the system when it is coupled to the harmonic oscillator
bath described by Eq.(2). The former is essentially rewriting Schrodinger equation in a quasi-classical
langauge and has nothing to do with the latter. Thus we write

\begin{eqnarray*}
\left(\frac{d W}{d t}\right) =  \left(\frac{\partial W}{\partial t}\right)_{dynamical} 
+\left(\frac{\partial W}{\partial t}\right)_{dissipative} \; \;.  
\end{eqnarray*}

While the dynamical evolution is governed by Wigner equation$^{(28)}$,

\begin{eqnarray*}
\left(\frac{\partial W}{\partial t}\right)_{dynamical} & = & \sum_{i=1}^N
\left[ -\frac{p_i}{2 m_i} \frac{\partial W}{\partial q_i}
+ \left( \frac{\partial V}{\partial q_i} \right) 
\frac{\partial W}{\partial p_i}
\right] \nonumber \\
& + & \sum_
{\begin{array}{c}
n_1+n_3+\cdots +n_N \; \; is \; \; odd \; \; and \; > 1 \\
\end{array}}
\left(\frac{\partial^{n_1+ \cdots n_N}V}{\partial q_1^{n_1} \cdots
\partial q_N^{n_N}} \right)  \frac{\left( \frac{\hbar}{2 i}\right)^
{n_1+ \cdots +n_N-1}}{n_1! \cdots n_N!}  \nonumber \\
& & \times \hspace{0.05cm} \frac{\partial^{n_1+ \cdots +N_N}}{\partial p_1^{n_1} \cdots \partial
p_N^{n_N}}W  \; \; ,
\end{eqnarray*}

the form of $\left(\frac{\partial W}{\partial t}\right)_{dissipative}$ is 
due to Caldeira and Leggett$^{(5)}$ as given by [when one of the system degrees of 
freedom is coupled to the reservoir as expressed in Hamiltonian (2); see the 
dissipative part of Eq.5.14 of Ref(5)]

\begin{eqnarray*}
\left(\frac{\partial W}{\partial t}\right)_{dissipative} = 2 \gamma 
\frac{\partial}{\partial p} p W + D \frac{\partial^2 W}{\partial p^2} 
\end{eqnarray*}

where $\gamma$ and $D$ have the same significance as in Eq.(3). The first term
in the last equation is a direct consequence of the existence of a
$\gamma$-dependent term in the imaginary part of the exponent in the expression
for the propagator for the density operator of Feynman and Vernon theory and 
has been
shown$^{(5)}$ to be responsible for appearance of a damping force in the classical
equation of motion for the Brownian particle to ensure quantum-classical correspondence.

The total dynamics is a superposion of two contributions provided by the last
two equations and when written elaborately we have;

\begin{eqnarray}
\frac{d W}{d t} & = & \sum_{i=1}^N
\left[ -\frac{p_i}{2 m_i} \frac{\partial W}{\partial q_i}
+ \left( \frac{\partial V}{\partial q_i} \right) 
\frac{\partial W}{\partial p_i}
\right] \nonumber \\
& + & \sum_
{\begin{array}{c}
n_1+n_3+\cdots +n_N  \; \; is \; \; odd \; and \; > 1 \\
\end{array}}
\left(\frac{\partial^{n_1+ \cdots n_N}V}{\partial q_1^{n_1} \cdots
\partial q_N^{n_N}} \right)  \frac{\left( \frac{\hbar}{2 i}\right)^
{n_1+ \cdots +n_N-1}}{n_1! \cdots n_N!}  \nonumber \\
& & \times \hspace{0.05cm} \frac{\partial^{n_1+ \cdots +N_N}}{\partial p_1^{n_1} \cdots \partial
p_N^{n_N}}W + 2 \gamma \frac{\partial}{\partial p} p W + D \frac{\partial^2 W}{\partial p^2} \; \;.
\end{eqnarray}

That the two contributions to the total evolution of the Wigner function in Eq.(4)
act independently in the overall dynamics is an assumption.
This assumption is also implicit in the operator master equation (3) and has been
routinely used in nonlinear and quantum optics, in general. (Note that Eq.(5.14)
of Ref(5) carry the same messege for a single-degree-of-freedom system).
Strictly speaking, the $\gamma$ and $D$ terms in Eqs.(3) and (4) are valid
if the system operators [i. e. , q and p in Eq.(2)] pertain to a harmonic oscillator.
When the system is nonlinear, as in the present case (also in many nonlinear 
optical situations) the usual practice is to add the additional contribution 
$-i [H_{non}, \rho]$ to the master equation [ in the language of Fokker-Planck
description this commutator, in general, contributes higher (third or more)
order derivatives of the distribution] and to assume that the dissipative terms
remain unaffected by the addition of the commutator term, $H_{non}$ being the 
nonlinear part of the Hamiltonian $H_{0}$. The validity of this assumption was
examined$^{(29)}$ earlier by Haake et. al. and also by us. It is now known that this
assumption is quite satisfactory within the
perview of weak damping and/ high temperature limit.

The equation (4) is a full quantum mechanical equation.
A simple version of the above equation for one-degree-of-freedom system was
used earlier by Zurek and Paz$^{(18)}$ for studying some interesting aspects of 
quantum-classical correspondence in relation to decoherence. The primary reasons for choosing 
Eq.(4) as our starting point for semiclassical analysis are (i) the rotating 
wave approximation (RWA) in the system-reservoir coupling has not been made
in deriving Eq.(4) (ii) Eq.(4) is also free from Born approximation (or weak-coupling 
approximation) ensuring that the theory is valid even in the strong damping 
limit. (iii) Eq.(4) reaches the correct
classical limit when $\hbar \rightarrow 0$, when D becomes the thermal 
diffusion coefficient in the high temperature limit. Thus Eq.(4) is a good 
description in the semiclassical limit.
Keeping in view of these  remarks and the earlier discussion in Sec IIA
we observe that Eq.(3) and Eq.(4) describe same dynamics in weakly dissipative 
systems. We adopt Eq.(4) for our semiclassical analysis that follows and Eq.(3)
for quantum numerical simulation in Sec IV to verify the theoretical propositions 
of this analysis.

The first term in Eq.(4) is the usual Poisson bracket which generates the Liouville
flow. Both the Poisson bracket and the 
higher derivative terms result from an
expansion of the Moyal bracket on the basis 
of an analytic $V(q)$. The last two 
terms have the same significance as in Eq.(3). It is important to note 
that the failure of correspondence between classical and quantum dynamics is 
predominantly due to higher derivative terms 
which make their presence felt
roughly beyond the Ehrenfest regime.

As a first step it is 
convenient to introduce the following scaling of c-numbers $\{ q_i, p_i \}$ in analogy to van Kampen's 
$\Omega$-expansion;

\begin{eqnarray}
q_i & = & q_i(t) + \hbar^{1/2} \mu_i \; \; \; , \nonumber\\
p_i & = & p_i(t) + \hbar^{1/2} \nu_i \; \; \; ,
\end{eqnarray}

where $\hbar$ is the associated 
smallness parameter for the present analysis.
$\mu$ and $\nu$ in Eq.(5) refer to quantum
fluctuations in  co-ordinate and momentum, 
respectively. $q(t)$ and $p(t)$ are the corresponding classical co-ordinate
and momentum. The time evolution of the distribution function of the 
fluctuation  variables obeys

\begin{equation}
\frac{\partial \phi (\{\mu_i\}, \{\nu_i\},t)}{\partial t}
= \sum_k \left[ - \frac{\nu_k}{m_k} \frac{\partial \phi}{\partial \mu_k} +
\mu_j \frac{\partial^2 V}{\partial q_j \partial q_k} \frac{\partial \phi}
{\partial \nu_k} \right]
+ 2 \gamma \frac{\partial}{\partial \nu} \nu \phi + {\cal O}(\hbar^{1/2}) \; \;.
\end{equation}

Although this equation does not involve any $\hbar$ explicitly, it describes the  
time evolution of probability density function $\phi( \left\{ \mu_i \right\}, 
\left\{ \nu_i\right\}, t)$
for the quantum noise variables $\left\{\mu_i, \nu_i \right\}$, since $\phi$  is the lowest
order quantum correction to classical distribution function $W\left( q_i(t), 
p_i(t), t \right)$. Secondly, the quantum dynamics enters into the picture when 
we put the quantum constraint(8) on the initial density function 
$\phi(\left\{\mu_k \right\}, \left\{ \nu_k \right\}, 0)$

\begin{equation}
\phi ( \left\{\mu_k \right\}, \left\{ \nu_k \right\}, 0) = \prod_{k=1}^N
\frac{1} {4 \sigma} \exp \left[ - \frac{\mu_k^2} {2 \sigma^2} - 2 \sigma^2 
\nu_k^2 \right] 
\end{equation}

\noindent
as

\begin{equation}
\langle (\Delta {\mu_i})^2 \rangle^\frac{1}{2}
\langle (\Delta {\nu_i})^2 \rangle^\frac{1}{2} 
=\sigma \cdot \frac{1}{2 \sigma} = \frac{1}{2} \; \;,
\end{equation}

\noindent
where $\hbar=1$ is used.

\noindent
We thus note that the initial density $\phi(\mu, \nu, 0)$ is not a 
$\delta$-function but has an appropriate spread. 
This spread incorporates the quantum noise which gets amplified as the density
$\phi$ evolves in time.
It is thus a quantum (minimum
uncertainty product) condition
and a requirement 
imposed by quantum-classical correspondence.
$\nu$ in Eq. 6 refers to the specific degrre of freedom of the system to which the reservoir is 
coupled to allow  the exchange of energy between the system and the reservoir.

As a second step we put Eq.(6)
in a more compact form by invoking the 
symplectic structure of the 
Hamiltonian dynamics. For this, we specify
\begin{equation}
z_i =
\left\{ \begin{array}{ll}
q_i & \; \; {\rm for} \; i=1 \cdots N  \; \;, \\
p_{i-N} & \; \; {\rm for} \; i=N+1, \cdots 2N \; \;.
\end{array}
\right.
\end{equation}

Defining I as 
\begin{equation}
I=
\left(\begin{array}{cc}
0 & E\\
-E & 0
\end{array}\right) \hspace{0.1cm},
\end{equation}
where E is an $N\otimes N$ unit matrix, one can write the Hamilton's equation
\begin{equation}
\dot{z}_i = \sum_j I_{ij} \frac{\partial H}{\partial z_j} \hspace{0.1cm}.
\end{equation}

Again we introduce the scaling $z_i$ as

\begin{equation}
z_i=z_i(t)+\hbar^{1/2} {\bf \eta}_i
\end{equation}

with

\begin{eqnarray}
\eta_i & = & \mu_i \,
\; \; \; \; \; \; {\rm for}\; \; i=1, \cdots N \nonumber \\
& = & \nu_{i-N} \; \; \; {\rm for}\; \; i=N+1 \cdots 2N  \; \;,
\end{eqnarray}

corresponding to quantum fluctuations in co-ordinates ($\mu_i$) and momenta ($\nu_i$).
We generalize Eq.(6) further to the extent that all the momentum components 
($\eta_i, \; \; i=N+1, \cdots 2N$) are coupled to the bath linearly.
One obtains the equation of motion for 
quantum fluctuation distribution function

\begin{equation}
\frac{\partial \phi}{\partial t} = - \sum_{i, j} \left[
J_{ij} \eta_i \frac{\partial \phi}{\partial \eta_j} - 2 \gamma_j
\frac{\partial}{\partial \eta_j} (\eta_j \phi) \right] \; \; \; ,
\end{equation}

where we have assumed that 
\begin{eqnarray*}
\begin{array}{lllll}
\gamma_j & = & 0 & \; \; \; \; {\rm for} \; \; j = 1, \cdots N \; \;, \\
\gamma_j & = & \gamma & \; \;\;{\rm for} \; \; j = N+1, \cdots 2N \; \;. 
\end{array}
\end{eqnarray*}

Here
\begin{equation}
J_{ij}= \sum_k I_{ik} \frac{\partial^2 H}{\partial z_k \partial z_j}
\end{equation}

contains the second derivatives of the potential and is a function of 
classical dynamical variables $z_i(t)$, (i. e., $p_i(t)$ and $q_i(t)$).

For further treatment Eq. (14) may be rewritten in a more compact form as 
follows:

\begin{equation}
\frac{\partial \phi}{\partial t} = [ -{\bf F}(t) \cdot {\bf \nabla} 
+ 2 N \gamma] \phi \; \; \; ,
\end{equation}
where
\begin{equation}
{\bf F}(t) = \underline{J}(t) {\bf \eta} - 2 \gamma {\underline K} {\bf \eta}  \; \; \; .
\end{equation}
${\bf \nabla}$ refers to 
differentiation with respect to the components of ${\bf \eta}$ and {\underline K}
is a $2N\otimes 2N$ matrix defined as
\begin{eqnarray*}
\begin{array}{lllll}
k_{ij} & = & 0 & \; \; \; {\rm for} \; \; i \ne j \; \;, \\
k_{ii} & = & 0 & \; \; \; {\rm for} \; \; i = 1, \cdots N  \; \;, \\
k_{ii} & = & 1 & \; \; \; {\rm for} \; \; i = N+1, \cdots 2N  \; \;. 
\end{array}
\end{eqnarray*}

\underline {$J$}, the jacobian matrix as defined in (15) is a function of classical dynamical variables 
$\{ q_i(t), p_i(t) \}$. The crucial question of stability/instability of classical
motion in Hamiltonian systems essentially rests on this jacobian, or curvature
(or second derivative) of the potential. Traditionally  the local linear stability
analysis around the fixed points is based on the assumption$^{(22,23)}$ of constant curvature.
However, the true stability of motion is only determined by keeping the time dependence of 
\underline {$J$} (implicitly through $\{q_i(t), p_i(t) \}$) matrix intact. Also
there is little connection between the local stability and global chaos. In 
view of this it is necessary to take full account of the time dependence
of the curvature of the potential \underline {$J$}(t) along the trajectory itself.
When the motion of the dynamical system is regular \underline {$J$}(t) is
highly correlated throughout the entire course of evolution. On the other
hand for chaotic motion when the dynamical variables in \underline {$J$}
(i. e. , $\{q_i(t), p_i(t) \}$) by virtue of the classical equation of motions for  $q_i(t)$
and $p_i(t)$ [or in general $z_i(t)$ of Eq.(9)] behaves {\it stochastically}
\underline {$J$}(t) describes a stochastic process. The loss of correlation
in chaotic dynamical systems thus rests on the decay of correlation of 
fluctuations of \underline {$J$}(t). What follows subsequently is a stochastic
description of classical chaos in terms of this correlation.

Ever since the early numerical study of Chirikov mapping$^{(30)}$ revealed that 
the motion of a phase space variable $\{q$ or $p\}$ can be characterized by a 
simple random walk diffusion equation, attempts have been made to describe
chaos in terms of a stochastic description ( Langevin and Fokker-Planck 
description has been widely employed). It has now been realized that 
deterministic maps can result in long time diffusional processes and methods 
have been developed to predict successfully the corresponding diffusion 
coefficients$^{(31)}$. In a number of recent studies$^{(11, 24, 25)}$ we have
shown that the fluctuation in the curvature of the potential is amenable to a
stochastic description in terms of the theory of multiplicative noise. This
allows us to realize a number of important results of nonequilibrium statistical
mechanics, like Kubo relation$^{(24)}$ fluctuation-decoherence 
relation$^{(25)}$ etc. in chaotic dynamics of a few-degree-of-freedom system.

Another important point to be noted here is that 
we {\it do not} make any {\it a priori assumption about the nature of the 
stochastic process} (${\bf J}(t)$). The special cases, such as, noise is 
Gaussian or $\delta$-correlated, etc.  have attracted so much 
attention in the literature
that it is necessary to emphasize that we have not made any such approximation. 
The stochasticity of ${\bf F}(t)$  
depends on \underline{J} which is determined by the exact solution of 
the classical equation of motion(11).
Eq.(16) may therefore be regarded as a stochastic 
differential equation with multiplicative noise. For convenience
${\bf F}(t) \cdot {\bf \nabla}$  can be 
partitioned (this partitioning will be clarified in more detail in the next section) into two parts; a constant part  ${\bf F}_0 \cdot \nabla$      
and a fluctuating part
${\bf F}_1(t) \cdot \nabla$. Thus we write
\begin{equation}
{\bf F} \cdot \nabla = {\bf F}_0 \cdot \nabla + {\bf F}_1 \cdot 
{\bf \nabla} \; \; \; .
\end{equation}

We now come to the  third step.
Making use of one of the main results for the theory of linear equation of 
the form (16) with multiplicative noise, we derive an average equation for 
$\phi$ as given by (for details, we refer to$^{(26,27)}$);

\begin{eqnarray}
\frac{\partial \langle \phi \rangle}{\partial t} & = & 
\left\{ -{\bf F}_0 \cdot \nabla
+ 2 N\gamma -\langle {\bf F}_1 \cdot \nabla \rangle + \int_0^\infty d\tau
\langle \langle {\bf F}_1(t) \cdot \nabla \exp ( -\tau [{\bf F}_0 \cdot \nabla
+2 N\gamma])  \right.\nonumber\\
& & \left.  {\bf F}_1(t-\tau) \cdot \nabla \rangle \rangle
\exp ( \tau [{\bf F}_0 \cdot \nabla + 2 N\gamma])  \right\} \langle \phi \rangle
\end{eqnarray}

where $\langle \langle \cdot \cdot \cdot \rangle \rangle $ implies
$\langle \langle q_i q_j \rangle \rangle 
= \langle q_i q_j \rangle -\langle q_i \rangle \langle q_j \rangle $.
The operator $\exp (-\tau {\bf F}_0 \cdot \nabla)$ provides the solution 
of the equation
\begin{equation}
\frac{\partial f(\eta, t)}{\partial t} = -{\bf F}_0 \cdot \nabla f(\eta,t)
\end{equation}

(where $f$ signifies the ``unperturbed'' part of $\phi$) which can be found 
explicitly in terms of characteristic curves. The equation
\begin{equation}
\dot{\eta} = {\bf F}_0 (\eta) 
\end{equation}
for fixed $t$ determines a unperturbed mapping from $\eta(\tau=0)$ to $\eta(\tau)$, i. e., 
$\eta \rightarrow \eta^\tau$ with inverse $(\eta^\tau)^{-\tau}=\eta$,
The solution of (20) is
\begin{equation}
f(\eta,t)= f(\eta^{-t}, 0) \left | \frac{d \eta^{-\tau}}{d \eta} \right | = 
\exp \left [ -t {\bf F}_0 \cdot \nabla \right ]  f(\eta,0 ).
\end{equation}
$\left | \frac{d\eta^{-t}}{d\eta} \right |$ being a Jacobian determinant. The
effect of $\exp(-t {\bf F}_0 \cdot \nabla)$ on $f(\eta)$ is as follows
\begin{equation}
\exp(-t {\bf F}_0 \cdot \nabla) f(\eta,0) = f(\eta^{-t}, 0) \left|\frac
{d \eta^{-t}}{d \eta} \right| \; \;.
\end{equation}

When this simplification is used in Eq.19 we obtain

\begin{eqnarray}
\frac{\partial \langle \phi \rangle}{\partial t} & = &
\left\{ -{\bf F}_0. {\bf \nabla}
+2 N\gamma - \langle {\bf F}_1.{\bf \nabla} \rangle +\int_0^\infty
d\tau \left| \frac{d{\bf \eta}^{-\tau}}{d{\bf \eta}}
\right| \right. \nonumber \\
& & \langle \langle {\bf F}_1({\bf \eta},t) 
.{\bf \nabla}_\tau {\bf F}_1({\bf \eta}^{-\tau}, t-\tau) \rangle 
\rangle \cdot \nabla_{-\tau} \left. \left| \frac{d\eta}{d \eta^{-\tau}}
\right|  \right\} \langle \phi \rangle \; \;.
\end{eqnarray}

The above consideration is based on  a second order expansion in   
$\alpha \tau_c$ (by van Kampen$^{(26)}$), where $\alpha$
is the strength parameter required for bookkeeping the order of the perturbation
fluctuation and $\tau_c$ is the correlation time of fluctuations in ${\bf F}_1(t)$
[in the derivation above we have put $\alpha = 1$]. 
The average $\langle \phi \rangle$ in Eq.(24) varies on a coarse-grained 
timescale which is much slower compared to the timescale set by the correlation time of fluctuation of 
${\bf F}_1(t)$. 
Second, the derivation above neglects the effects of 
higher powers of $\hbar$ and thus the Eq.(24) is an effective 
semiclassical  equation for quantum fluctuation distribution function. Since 
it contains second derivatives with respect to components of   
$\eta$, it has the form of a 
Fokker-Planck equation. Third, the theory discussed so far (Eq.(24))
is valid, in general,
for N-degree-of-freedom systems.

We now adapt Eq.(24) to the classic paradigm of chaotic dynamics  -  the 
Henon-Heiles system.

\section{The dissipative Henon-Heiles system}
\vspace{0.25cm}

\subsection{The Fokker-Planck equation}
\vspace{0.25cm}

We consider the
Henon-Heiles system which is kept in contact with the surroundings. The Hamiltanian of
this system is given by

\begin{equation}
H_0 = \frac{p_{1}^2}{2 } +\frac{p_{2}^2}{2} + V(q_{1},q_{2})   \; \; ,
\end{equation}

where $V(q_1, q_2)= \frac{1}{2} (q_1^2+q_2^2+2q_1^2q_2-\frac{2}{3}q_2^3)$, 
is the potential energy of the two-degree-of-freedom system.

The classical equations of motion of the particle in presence of damping (at 
a rate $\gamma$) are

\begin{eqnarray}
\dot{q_{i}} & = & p_{i} \; \; \; ,  \nonumber \\
\dot{p_{i}} & = & 
- \gamma p_{i} -\frac{\partial V(q_{1},q_{2})}{\partial q_{i}}  \; \; \; ,  i=1,2 \; \; \;.
\end{eqnarray}

Note that in the above equation we have assumed for simplicity the value of 
dissipation rate same for both the degrees of freedom.
The equations of motion for the quantum fluctuation variables  
$\eta_1$ , $\eta_2$, $\eta_3$ and $\eta_4$ corresponding
to $q_{1}$, $q_{2}$ , $p_{1}$ and $p_{2}$, respectively, read as follows:

\begin{equation}
\frac{d}{dt} \left( 
\begin{array}{c}
\eta_1\\
\eta_2 \\
\eta_3 \\
\eta_4
\end{array} \right) =
\underline{J} 
\left(
\begin{array}{l}
\eta_1\\
\eta_2\\
\eta_3\\
\eta_4
\end{array}
\right)    .
\end{equation}

Following the procedure as described in the last section    
$\underline{J}$ can be identified as

\begin{equation}
\underline{J} = \left(
\begin{array}{ccccccc}
0 & \; & 0 & \; & 1 & \;  & 0 \\
0 & \; & 0 & \; & 0 & \; & 1 \\
-1-\zeta_1(t) & \; & \zeta_2(t) & \; & 0 & \; & 0 \\
\zeta_2(t) & \; & -1+\zeta_1(t) & \; & 0 & \; & 0
\end{array}
\right) \; \; \; .
\end{equation}

Here  $\zeta_{1}(t)$ and $\zeta_{2}$ (t) are given by
\begin{eqnarray}
\zeta_{1}(t) = 2 q_{2}  \; \; \;,  \nonumber\\ 
\zeta_{2}(t)=-2 q_{1} \; \; \; .
\end{eqnarray}

Since both $q_1, q_2$ are determined by classical equations of motion (26), 
chaoticity of the trajectory imparts stochasticity in the dynamics of quantum 
fluctuations in Eq.(27). Thus, as elaborated in the last section, $\zeta$(t)
terms represent the stochastic part of the second derivative of the potential
$V(q_1, q_2)$.

If one takes into consideration of the     
$\gamma$-term then ${\bf F}(t)$ in Eq.(18) can be 
written as,

\begin{equation}
{\bf{F}}={\bf{F}}_0+{\bf{F}}_1(t) \; \; \; ,
\end{equation}
where
\begin{equation}
{\bf{F}}_0  =  \left(
\begin{array}{c}
\eta_3 \\
\eta_4 \\
-\eta_1-2\gamma \eta_3 \\
-\eta_2-2\gamma \eta_4
\end{array}
\right)
\end{equation}
and
\begin{equation}
 {\bf{F}}_1(t) = \left(
\begin{array}{c}
0 \\
0 \\
-\zeta_1(t)\eta_1+\zeta_2(t)\eta_2 \\
\zeta_2(t)\eta_1 + \zeta_1(t)\eta_2
\end{array}
\right)  .
\end{equation}

The equations for the characteristic curves are
\begin{eqnarray}
\dot{\eta}_1 & = &  \eta_3 \; \; \; , \nonumber \\
\dot{\eta}_2 & = &  \eta_4  \; \; \; , \nonumber \\
\dot{\eta}_3 & = &  
- \zeta_1(t) \eta_1 + \zeta_2(t) \eta_2 - 2\gamma \eta_3 
- \eta_1 \; \; \; , \nonumber \\
\dot{\eta}_4 &= & \zeta_2(t) \eta_1 + \zeta_1(t) \eta_2 - 2\gamma \eta_4 
- \eta_2 \; \; \; .     
\end{eqnarray}

Eq.(33) describes the dynamics of quantum fluctuations in presence of dissipation.
The mapping $\eta \rightarrow \eta^t$        
can be found by solving the unperturbed version of 
Eq.(33) (i. e. the $\zeta$ terms are omitted) for discrete small steps of 
$\tau$ (which is consistent with the requirement that the correlation time
is short and finite) and is given by [we refer to van Kampen$^{(26)}$ for
details of treatment of multiplicative stochastic noise in Eq.(33)]

\begin{eqnarray}
\eta_1^{-\tau} & = & -\tau\eta_3+\eta_1 \; \; \;, \nonumber \\
\eta_2^{-\tau} & = & -\tau\eta_4+\eta_2 \; \; \;, \nonumber \\
\eta_3^{-\tau} & = & \frac{\eta_1}{2\gamma}(e^{2\gamma\tau}-1)
+\eta_3e^{2\gamma\tau} \; \; \;, \nonumber \\
\eta_4^{-\tau} & = & \frac{\eta_2}{2\gamma}(e^{2\gamma\tau}-1)
+\eta_4e^{-2\gamma\tau} \; \; .
\end{eqnarray}

The Jacobian determinant of the transformation reads as

\begin{equation}
\left| \frac{d \eta^{-\tau}}{d \eta} \right| 
\simeq e^{4\gamma \tau} \; \;,
\end{equation}

where the terms of the order of $\tau^2$ are neglected. This is well 
within the error bound $(\alpha^2 \tau_c)$ as shown by van Kampen$^{(26)}$ and 
does not incorporate additional error in the analysis.

Also note that
\begin{equation}
\left| \frac{d \eta}{d \eta^{-\tau}} \right|
\simeq e^{-4\gamma \tau} .  
\end{equation}

Making use of the mapping transformations             
$\eta \rightarrow \eta^\tau$ (Eq.34) one 
calculates the first and second derivative terms in Eq.(24) [For details we 
refer to Ref(26)]. 
The master equation (24) for the Henon-Heiles
system can then be written down. This is
\begin{eqnarray}
\frac{\partial \langle \phi \rangle}{\partial t}
& = & \left[ 
-\eta_3 \frac{\partial}{\partial \eta_1} - \eta_4 
\frac{\partial}{\partial \eta_2} + \{ \eta_1 +2\gamma \eta_3 -
(C_3 \eta_1-B_3\eta_2+A_3\eta_2 \right. \nonumber \\
& - & B_3'\eta_1+A_3\eta_1+B_3'\eta_2+B_3\eta_1+C_3\eta_2)-
(\langle \zeta_2(t) \rangle \eta_2 - \langle \zeta_1(t) \rangle \eta_1)\}
\frac{\partial}{\partial \eta_3} \nonumber \\
& + & \{\eta_2+2\gamma\eta_4-(\langle \zeta_2 \rangle \eta_1+\langle\zeta_1
\rangle \eta_2)+(B_3\eta_2-C_3\eta_1-A_3\eta_2-B_3'\eta_1 \nonumber \\
& - & A_3\eta_1-B_3'\eta_2+B_3\eta_1-C_3\eta_2)\}
\frac{\partial}{\partial\eta_4}+4 \gamma
+E_1\frac{\partial^2}{\partial \eta_3 \partial\eta_1}
+E_2\frac{\partial^2}{\partial \eta_4 \partial\eta_2} \nonumber \\
& + & F_1\frac{\partial^2}{\partial\eta_3\partial\eta_2} +
F_2\frac{\partial^2}{\partial\eta_4\partial\eta_1}+
G(\frac{\partial^2}{\partial\eta_3^2}+\frac{\partial^2}{\partial\eta_3
\partial\eta_4}) \nonumber \\
& + &
\left. H(\frac{\partial^2}{\partial\eta_4 \eta_3}
+\frac{\partial^2}{\partial\eta_4^2}) \right] \langle \phi \rangle \; \;,
\end{eqnarray}
where
\begin{eqnarray}
E_1 & = & A_4\eta_2^2+C_4\eta_1^2-\eta_1\eta_2(B_4'+B_4) \; \; ,\nonumber \\
E_2 & = & A_4\eta_1^2+\eta_1\eta_2(B_4+B_4')+C_4\eta_2^2 \; \; ,\nonumber \\
F_1 & = & \eta_1\eta_2(A_4-C_4)+B_4\eta_2^2-B_4'\eta_1^2 \; \; ,\nonumber \\
F_2 & = & \eta_1\eta_2(A_4-C_4)-B_4\eta_1^2+B_4'\eta_2^2 \; \; ,\nonumber \\
G & = & \eta_2^2(A_2+B_2)+\eta_1^2(C_2-B_2')+\eta_1\eta_2(A_2-C_2
-B_2'-B_2) \nonumber \\
& & + \eta_1\eta_4(B_3'+C_3)+\eta_2\eta_3(B_3-A_3)+\eta_1\eta_3(B_3'-C_3)
-\eta_2\eta_4(A_3+B_3) \; \; , \nonumber \\
H & = & \eta_2^2(B_2'+C_2)+\eta_1^2(A_2-B_2)+\eta_1\eta_2(A_2-C_2
+B_2+B_2')  \nonumber \\
& & - \eta_1\eta_4(A_3+B_3)+\eta_1\eta_3(B_3-A_3)-\eta_2\eta_4(B_3'-C_3)
\nonumber \\
& & + \eta_2\eta_3(C_3-B_3') \; \;,
\end{eqnarray}
and
\begin{equation}
\left. \begin{array}{ccccccc}
A_2 & = & \int_0^\infty \langle \langle \zeta_2(t) \zeta_2(t-\tau)
\rangle \rangle e^{-2\gamma \tau} d\tau & , &
B_2' & = & \int_0^\infty \langle \langle \zeta_1(t) \zeta_2(t-\tau)
\rangle \rangle e^{-2\gamma \tau} d\tau \\
& & & & & & \\
A_3 & = & \int_0^\infty \langle \langle \zeta_2(t) \zeta_2(t-\tau)
\rangle \rangle e^{-2\gamma \tau} \tau d\tau & , &
B_3' & = & \int_0^\infty \langle \langle \zeta_1(t) \zeta_2(t-\tau)
\rangle \rangle e^{-2\gamma \tau} \tau d\tau \\
& & & & & & \\
A_4 & = & \int_0^\infty \langle \langle \zeta_2(t) \zeta_2(t-\tau)
\rangle \rangle \tau d\tau & , &
B_4' & = & \int_0^\infty \langle \langle \zeta_1(t) \zeta_2(t-\tau)
\rangle \rangle \tau d\tau \\
& & & & & & \\
B_2 & = & \int_0^\infty \langle \langle \zeta_2(t) \zeta_1(t-\tau)
\rangle \rangle e^{-2\gamma \tau} d\tau & , &
C_2 & = & \int_0^\infty \langle \langle \zeta_1(t) \zeta_1(t-\tau)
\rangle \rangle e^{-2\gamma \tau} d\tau \\
& & & & & & \\
B_3 & = & \int_0^\infty \langle \langle \zeta_2(t) \zeta_1(t-\tau)
\rangle \rangle e^{-2\gamma \tau} \tau d\tau & , &
C_3 & = & \int_0^\infty \langle \langle \zeta_1(t) \zeta_1(t-\tau)
\rangle \rangle e^{-2\gamma \tau} \tau d\tau  \\
& & & & & & \\
B_4 & = & \int_0^\infty \langle \langle \zeta_2(t) \zeta_1(t-\tau)
\rangle \rangle \tau d\tau & , &
C_4 & = & \int_0^\infty \langle \langle \zeta_1(t) \zeta_1(t-\tau)
\rangle \rangle \tau d\tau 
\end{array} \right \} \; \;.
\end{equation}

The above equation (37) is a Fokker-Planck equation for probability 
distribution  of quantum fluctuations for the 
dissipative Henon-Heiles system.
It is evident that stochastic averaging over classical chaos leads to the 
average equation and the correlation functions contained in $A_2,...., C_4$ . 
The correlation of 
fluctuations of curvature of the classical potential thus determines the 
drift 
and diffusion terms of the Fokker-Planck equation. It must be emphasized
that this fluctuation has nothing to do with the stochasticity inherent in
the system-heat bath model governed by Hamiltonian(2). We also point out that
since the very notion of stochastic process in describing the curvature of the
potential results in the diffusion terms, the stochastification imparts a kind
of irreversibility in the evolution governed by the Fokker-Planck equation(37).
The origin of this irreversibility is classical chaos and not due to any 
external influence. This is characteristic of the nonlinear system, itself.

\vspace{0.25cm}

\subsection{The solution of Fokker-Planck equation}
\vspace{0.25cm}

The appearance of
the variables $\eta_1, \eta_2, \eta_3$ and $\eta_4$
in the diffusion terms precludes the possibility of an exact 
solution of Eq.(37). One thus takes resort to {\it weak noise approximation} 
(this is consistent with assumption that fluctuations are not too large)
scheme. The diffusion terms (given in the Appendix) are  thus assumed to be 
constant. 
 
The resulting Fokker-Planck equation can be transformed to the following 
simple form

\begin{equation}
\frac{\partial \langle \phi \rangle}{\partial t}
 =  [ \lambda u \frac{\partial}{\partial u}+ A \frac{\partial^2}{\partial u^2} 
 +4 \gamma ] \langle \phi \rangle
\end{equation}

where

\begin{equation}
u = a \eta_1+b \eta_2 +c \eta_3 +\eta_4 \; \; ,
\end{equation}

\noindent
and the constants $\lambda$, A, a, b and c are given in the appendix.

We then search for the Green's function or conditional probability solution 
for the system at $u$ at time $t$ given that it had the 
value $u'$ at $t=0$. The initial condition which is 
required to bring forth quantum-classical correspondence is represented by

\begin{equation}
p(u,t=0) = 
\frac{\epsilon}{\pi} e^
{-\epsilon (u-u')^2 } \; \;.
\end{equation}

This means that $\epsilon$ should be chosen in such a way that corresponds to
the minimum uncertainty product of the initial wave packet. For notational 
convenience we have used

\begin{equation}
\langle \phi(u,t) \rangle =  p(u,t) \; \;.
\end{equation}

We now look for a solution of the equation (40) of the form
\begin{equation}
p(u, t) | u',0) = e^{G(t)},
\end{equation}
where
\begin{equation}
G(t)=-\frac{1}{\Gamma(t)}(u-\Omega(t))^2
+{\rm ln}\nu(t) \; \; .
\end{equation}

We are to see that, we can, by suitable choice of
$\Omega(t)$, $\Gamma(t)$ and $\nu(t)$, 
solve Eq.(40) subject to the initial condition

\begin{equation}
p(u, 0) | u', 0) = 
\frac{\epsilon}{\pi} e^
{-\epsilon (u-u')^2 } \; \; \; .
\end{equation}

Comparison of this with (44) with G(0) shows that
\begin{equation}
\Gamma(0) = \frac{1}{\epsilon} \; \; , \; \; 
\Omega(0) = u' \; \; , \; \;
\nu(0)=\frac{\epsilon}{\pi} \; \; .
\end{equation}

If we put (44) in (40) and equate the coefficients of equal powers of $u$
we obtain after some algebra the  following 
set of equations
\begin{equation}
\frac{1}{\Gamma^2} \frac{d \Gamma}{dt} = -\frac{\gamma'}
{\Gamma} +\frac{D_1}{\Gamma^2} \; \; , 
\end{equation}
\begin{equation}
\frac{d\Omega}{dt}  =  -\lambda \Omega  \; \;,
\end{equation}
and
\begin{equation}
\frac{1}{\nu} \frac{d\nu}{dt} = 4\gamma -\frac{D_{1}}{2\Gamma}  \; \; \;,
\end{equation}
where
\begin{eqnarray}
\gamma' & = &2\lambda  \; \; \;, \nonumber \\
D_1 & = &4A  \; \; \;.
\end{eqnarray}

The relevant solution of $\Gamma(t)$ for the present problem which
satisfies the initial conditions above is given by
\begin{equation}
\Gamma(t) = \Gamma(0)e^{-\gamma' t}+\frac{D_{1}}{\gamma'}(1-e^{-\gamma' t}) \; \;.
\end{equation}

It is important to note that the expansion of the wave packet is determined
by $\Gamma$(t) which is controlled by the two parameters, $D_1$ and $\gamma'$
which by the virtue of Eqs.(51) and (40) can be identified as the 
``renormalized'' diffusion and drift coefficients, respectively. The origin of this
``renormalization'' is essentially classical chaos since these coefficients
are the complicated functions of the correlation function of the 
fluctuations of the curvature of the potential.

\subsection{Results: quantum fluctuations, expansion of phase space and entropy}
\vspace{0.5cm}

Having obtained $p(\eta_1, \eta_2, \eta_3, \eta_4)$ 
we are now in a position 
to determine the various theoretical quantities. We calculate the quantum 
fluctuations of position and momentum 
variables. Since the conditional probability $p$ is given 
is given 
by Eq.(44), this
together with (48-50) may be employed to calculate first and second
moments. Thus we express
\begin{equation}
\langle \eta_1 \rangle = \frac{
{\int \int \int \int}_{-\infty}^\infty
p(\eta_1, \eta_2, \eta_3, \eta_4, t | \eta'_1, \eta'_2, \eta'_3, \eta'_4, 0)
\eta_1 d\eta_1 d\eta_2 d\eta_3 d\eta_4}
{{\int \int \int \int}_{-\infty}^\infty
p(\eta_1, \eta_2, \eta_3 ,\eta_4, t | \eta'_1, \eta'_2, \eta'_3 \eta'_4, 0)
d\eta_1 d\eta_2 d\eta_3 d\eta_4}
\end{equation}

in terms of conditional probability $p$. Explicit 
calculation yields
\begin{equation}
\langle \eta_1 \rangle = \frac{\Omega(t)}{a} \; \;.
\end{equation}
 Similarly we obtain
\begin{equation}
\langle \eta_1^2 \rangle = \frac{1}{2a^2} \Gamma(t) + \frac{{\Omega(t)}^2}{a^2} \; \;.
\end{equation}
The conjugate variable to $\eta_{1}$ is $\eta_{3}$ whose 
average is given by
\begin{equation}
\langle \eta_3 \rangle = \frac{\Omega(t)}{c} \; \;.
\end{equation}
Similarly
\begin{equation}
\langle \eta_3^2 \rangle = \frac{1}{2c^2} \Gamma(t)+ \frac{{\Omega(t)}^2}{c^2} \; \; \;.
\end{equation}

Therefore the uncertainty in coordinate     
$\Delta \eta_1$ and that in its conjugate momentum $\Delta \eta_3$      
are  obtained as follows;
\begin{eqnarray}
\Delta \eta_1^2 & = & \langle \eta_1^2 \rangle - \langle \eta_1 \rangle^2
=\frac{1}{a^2} \left[ \frac{\Gamma(t)}{2} \right]  \; \; \; ,\\
\Delta \eta_3^2 & = & \langle \eta_3^2 \rangle - \langle \eta_3 \rangle^2
= \left[ \frac{\Gamma(t)}{2} \right] \frac{1}{c^2} \; \; \; ,
\end{eqnarray}

where the relations (54-57) have been used. The uncertainty product
$\Delta \eta_1 \Delta \eta_3$ at any time is given by
\begin{equation}
\Delta \eta_1 \Delta \eta_3 = \frac{1}{2|a|c} \Gamma(t)  \; \; \;,
\end{equation}

\noindent
where $\Gamma(t)$ is determined by Eq.(52) 
subject to  initial conditions (47). This implies that we are to choose 
$\epsilon=\frac{1}{|a|c}$ to satisfy the minimum uncertainty 
product condition for t=0, for the wave packet [i.e., 
$\Delta \eta_1 \Delta \eta_3 = \frac{1}{2} $ ] .

We now discuss the following results ;

(i) The  relation (60) illustrates the evolution of 
quantum fluctuation as a function of time in terms of 
$\Gamma(t)$ which by the 
virtue of Eq.(52) is determined by the initial condition $\Gamma(0)$ 
[Eq.(47)]
and the other two parameters $D_1$ and $\gamma'$. {\it The early expansion of 
quantum fluctuations} has been recognized {\it as a typical signature of classical
chaos} on a generic quantum phenomenon$^{(24,25)}$. Note that $D_1 [ = 4A$, see Eq.(51)]
is the diffusion 
coefficient that appeared in the Fokker Planck Eq.(40)
[this is not to be confused with the thermal
diffusion coefficient $D$ in Eq.(4) which arises due to the
interaction with the
surroundings] and $\gamma'$ refers to the modified dissipation rate of the system   
in contact with the surroundings and is related to 
$\lambda$ [by Eq.(51)] which is determined by Eq.(A13). The diffusion
coefficient $D_1$ and modification of
dissipation rate are due to the correlation of fluctuations of the 
curvature of the classical potential $\zeta_1(t)$ and $\zeta_2(t)$
through $A_{2} ......C_{4}$ in Eqs.(39) and (A1).  
The origin of diffusion coefficient $D_1$ and the modification of $\gamma$
thus have purely deterministic origin.

To analyze the growth of quantum fluctuations quantitatively 
[Eq.60] we first 
consider the dissipative classical chaotic motion governed by Eq.(26). We
choose the initial conditions for energies 
$\frac{1}{8}$ and $\frac{1}{6}$. These energy values are 
wellknown in the context of classical Henon-Heiles Hamiltonian. 
It is important to note that even within the restricted domain of
weak dissipation, the dissipative Henon-Heiles system approaches a manifold 
reduced dimensionality. The timescale over which this reduction takes place
is determined essentially by the magnitude of the damping constant $\gamma$.
This classical behaviour is illustrared in Figs. 1 $(a)$ and 2 $(a)$ for $\gamma= 0.001$
for the energies $\frac{1}{8}$ and $\frac{1}{6}$, respectively. Figs. 1 $(b)$
and 2 $(b)$ depict the corresponding Poincare maps for the conservative 
$(\gamma = 0.0)$ Henon-Heiles system. It is thus 
apparent that even weak dissipation profoundly alters the characteristics of the
stochastic process represented by the classical Hamiltonian chaos.
The attractor clearly lies at the center.
To calculate classical ensemble average of the quantities like     
$\langle \zeta(t) \rangle$ and $ \langle \langle \zeta(t) \zeta(t-\tau) 
\rangle \rangle$,
we carry out averaging over long time series for the given initial condition. 
The numerical procedure has been discussed earlier in Ref$^{(24)}$.

Following Eq.(60) we plot the variation of uncertainty product
$ [\Delta \eta_1 \Delta \eta_3]$ ($\Delta \eta_1$ and $\Delta \eta_3$              
are the quantum variances 
corresponding to position and momentum for one degree of freedom, respectively) as a 
function of time for different values of energy corresponding to chaotic 
trajectories (damping rate $\gamma =0.001$)
in Figs 3 and 4. It has already been pointed out 
that the major input for the theoretical 
quantity are the chaotic diffusion coefficient $D_{1}$ and $\gamma'$ which 
are further related to 
$A_{2}$ ..... $C_{4}$ , i.e., to classical correlation functions of the 
curvature of the potential. The theoretical curves are 
denoted in 
Figs 3 and 4 by the dotted lines. 

(ii) The relation (60) also shows that there exist {\it a critical limit to the
expansion of phase space}.
This limit is given by

\begin{equation}
\Delta \eta_1 \Delta \eta_3 |_{t \rightarrow \infty}   = \frac{D_1}{2|a|c \gamma'} \; \;. 
\end{equation}

The existence of this critical width is a consequence of the competetion 
between chaotic diffusion, which attempts to expand the wave packet and 
dissipation $\gamma$ which has the opposite tendency and this interplay ultimately
leads to a compromised steady state.

At this juncture it is necessary to clarify the concept of steady state of the
quantum dissipative system as applied here. The Henon-Heiles Hamiltonian is
strongly nonlinear and so the notion of the thermodynamic equlibrium is 
inappropriate. Instead we mean a stationary state of the quantum system in the
following sense. Here we are concerned with an asymptotic distribution
of quantum noise variables $\eta_i$ [$z_i= z_i(t)  + \hbar^\frac{1}{2} \eta_i$,
$z_i(t)$ being the classical position or momentum variable] in terms of the
probability distribution function $\langle\phi(u, t) \rangle$ [=p(u, t), u
being the combination of quantum noise variables $\eta_i$ (see Eq. 41)],
rather than a distribution of $z_i(t)$-s. Note that this function does not 
involve any classical contribution
$z_i(t)$ directly. The classical chaotic fluctuation in $z_i(t)$-s contribute to $\eta_i$ via their classical 
correlation functions (in a, b, c of Eq. 41). 
These correlations primarily determine the dynamics of $\eta_i$ [see (49) and
(54)] through $\lambda$ where the role of $\gamma$ is not the dominant one.
This implies that although the 
classical motion of $z_i(t)$ settles down on an attractor approximately on a
scale of, say, $\gamma^{-1}$ as noted in Figs.1 $(a)$ and 2 $(a)$, the quantum noise 
$\eta_i$-s approach the steady state very rapidly, i. e., in a few time units.
Thus the present quantum steady state does not correspond to a settling down
of classical motion on a attractor. Had we consider the asymptotic 
distribution of c-number variables $z_i$ through the Wigner density 
function W($z_i, t$) ,
then that would have correspond to an ideal quantum stationary state.

(iii) To make our analysis of irreversible evolution in presence of classical 
chaotic diffusion more quantitative, it is useful to calculate the entropy
S of the Gaussian state by defining it$^{(2)}$ as

\begin{equation}
S = - p(t) {\ln} p(t)
\end{equation}
where p is as defined by Eq.44.
In Fig 5 we show the evolution of entropy due to quantum noise corresponding 
to the classical trajectories of the dissipative Henon-Heiles system for the
energies $E= \frac{1}{6}$ and $\frac{1}{8}$ and $\gamma = 0.001$. At a very early stage the entropy 
change remains very small. It is then followed by a sharp increase and then
finally tends to increase at a very slow rate. It is interesting to note
that Zurek and Paz$^{(18)}$ advocated the efficacy of studying the evolution 
of entropy as a conseqence of interplay between Liouville dynamics and high
temperature surrounding to examine the hall-mark of a nonintegrable system.
Similar attempts had been made by us$^{(11)}$ earlier using Husimi distribution function
to identify the different stages of quantum evolution.

Since the theory of stochastic fluctuations of the curvature of the potential
rests on  van Kampen's expansion in $\alpha \tau_c$ as emphasized earlier, care 
should be taken to calculate the integrals (39) over the correlation functions. 
To implement this numerically one
considers the first fall of the correlation functions to adjust the cut off
in time for numerical evalution of the integrals. This is a crucial 
requirement for the theory which should be appropriately taken care of 
in numerical calculation.

We now point out a pertinent issue in the context of the nature of  
classical stochastic process considered here. It is wellknown that the
Henon-Heiles model (without dissipation) is a typical KAM problem, i. e., it
represents  {\it soft chaos} and in principle never reaches the limit of 
fully developed {\it hard chaos}. This implies that, in a sense, classical
noise has, in general, not very short correlation time. However the systematic
procedure to deal with long correlation time when the nature of noise
is rather unknown is relatively scarce. In principle, van Kampen's strategy as 
adopted here is applicable for consideration of higher order non-Markovian
contributions. But such
an extension is rather complicated both from analytical and numerical point of view.
We therefore confine ourselves to the lowest order non-Markovian contribution
to noise arising out of classical stochasticity and point out that such 
a description is not inappropriate in view of the short timescale over which
the quantum fluctuations $\eta_i$(t) determined by the correlations of the 
classical noise, persist.

\vspace{0.5cm}

\section{Numerical simulation of the quantum operator master equation}

\vspace{0.5cm}

For a full quantum-mechanical calculation to verify the basic 
theoretical propositions of semiclassical dynamics, we now return to Eq.(3).
To solve the Eq.(3) for the Henon-Heiles system we choose two sets of basis 
vectors
$\{| n_{1} \rangle \}$ and $\{|n_{2} \rangle \}$ of two different harmonic 
oscillators which satisfy 
$(\hat{p_{1}}^2/2m_{1} + (1/2) m_{1} \omega_{1}^2 \hat{q_{1}}^2) | n_{1} \rangle 
=[(n_{1}+1/2)\hbar \omega_{1}] | n_{1} \rangle$  and
$(\hat{p_{2}}^2/2m_{2} + (1/2) m_{2} \omega_{2}^2 \hat{q_{2}}^2) | n_{2} \rangle 
=[(n_{2}+1/2)\hbar \omega_{2}] | n_{2} \rangle$ . The 
frequencies $\omega_{1}$ and $\omega_{2}$ are arbitrarily 
adjusted to economize the size of the basis 
set. For the present purpose we choose  
$\omega_{1}=6.25$, $\omega_{2}=6.20, \hbar=1$, and 35
basis vectors.

Quantum-classical correspondence 
is maintained through the construction of minimum
uncertainty wave packets         
$| \alpha_{q_{i}, p_{i}} \rangle$ of Gaussian form in position and momentum 
representations having position    
$\langle q_{i} \rangle$ and average momentum     
$\langle p_{i} \rangle$ such that
\begin{equation}
\langle \alpha_{q_{i}, p_{i}} | n_{i} \rangle = \left[ \exp(-0.5 | \alpha_{i} |^2 )
\right] \frac{\alpha_{i}^n}{\sqrt{n_{i}!}} \; \; \; , 
\end{equation}
where,
\begin{eqnarray*}
\alpha_{i} = \sqrt{m_{i} \omega_{i}/2} [ \langle q_{i} \rangle + (i/m_{i}\omega_{i}) \langle p_{i} 
\rangle ]  \; \; \; ,  i=1,2 \; \; \; .
\end{eqnarray*}

The quantum evolution is followed by locating the average positions and average
momenta of the initial wave packets corresponding to the initial positions and 
momenta of two classically chaotic trajectories for the two  energy values
$\frac{1}{8}$ and $\frac{1}{6}$.  
Another important check for the 
numerical 
calculation is to keep the trace of the density matrix [Eq.(3)] equal to unity 
for the entire evolution.
We have also checked that the result is robust against the variation of the
size of the basis set.
The numerical curves (solid lines) have been superimposed in Figs 3 and 4 for the corresponding
values of energy.
It may be observed that the agreement between
the theoretical and numerical curves is quite satisfactory. This justifies
the validity of our semiclassical approach.

\vspace{0.5cm}

\section{Discussion on the approximations, summary and conclusions}

\vspace{0.5cm}

Based on a traditional scheme of system-reservoir model we have developed a 
theory of dissipative chaotic system. We make use of appropriate 
$\hbar$-scaling analogous to van Kampen's $\Omega$-expansion, of equation for
Wigner quasi-probability distribution functions which takes into account of
thermal diffusion and dissipation due to the reservoir. We have shown that the
semiclassical approximation leads us to an equation of motion for Wigner
function for quantum noise which is governed by the dissipation due to reservoir and
the second derivative of the classical potential, latter being a key-point
in determining the stability of classical motion. Since chaoticity 
originates from the exponential loss of correlation of initially nearby
trajectories this derivative behaves as a 
stochastic (deterministic) process. This stochastic process is amenable to
a theoretical analysis (without imposing any a priori assumption about its 
nature) in terms of a treatment of stochastic differential equation with 
multiplicative noise. The resulting Fokker-Planck equation carries the 
information about the drift and diffusion coefficients which are expressible in terms
of correlation functions of fluctuations of the curvature of the classical
potential. As a prototypical example we have illustrated our analysis with 
the help of the Henon-Heiles Hamiltonian.

We now make a few remarks on the approximations involved in the  present 
treatment and their validity.

(i) It must be emphasized that since the system-reservoir dissipative dynamics
as governed by the operator master equation (3) is based on Born-Markov approiximation (the 
correlation time of the reservoir must be very short (Markov) for the
interaction between the system and the reservoir to be sufficiently small
(Born/weak coupling)), the underlying stochastic process due to the reservoir
is {\it Markovian by construction}. On the other 
hand the stochasticity due 
to classical chaos as inherent in the 
fluctuations of the curvature of the potential
is {\it non-Markovian} since we take account of short but finite 
correlation time of this fluctuations. The construction of the associated 
Fokker-Planck equation is based on a perturbative cumulant expansion in 
$\alpha \tau_c$, where $\tau_c$ is the correlation time of fluctuation of 
the curvature of the potential. 
The convergence of expansion as demonstrated by van Kampen$^{(27)}$ thus allows
us to retain only upto second derivative terms and as such one need not go to
third or higher order terms to describe the dynamics.
While we note that there exist a vast body of literature in condensed matter 
and chemical physics dealing with finite time response of the reservoir
which results in frequency dependence of friction coefficient $\gamma$, these
and related aspects of dissipative dynamics are outside the scope of the
present treatment. 
Our approach is similar to Graham et. al$^{(7)}$  and Milonni et. al$^{(8)}$ in this respect.
Thus the
short time regime, we believe, is mainly controlled by the curvature of the
potential and the correlation of its 
fluctuations or in other words the short time dynamics is dominated by 
characteristic motion of the system itself.
However, in the ultimate passage towards equilibrium the dissipation plays 
a prominent decisive role.

(ii) In the master equation(4), the classical stochasticity due to chaos and
the quantum noise due to incoherent processes induced by the heat bath act 
simultaneously and influence one another. We have already noted that because
of $\hbar$-scaling of this equation, one arrives at Eq.6 in which the quantum
noise term D due to surrounding does not appear in the lowest order. 
Thus it is because of strict
semiclassical nature of our approach which is consistent with the 
consideration of dissipative contribution of 
$(\frac{\partial W}{\partial t})_{dissipative}$ in Eq.(4) 
(valid for $kT > \hbar \omega$, because the propagator in the 
relevant integral form of the density operator master equation of Leggett and
Caldeira [5.14 of Ref(5)] has been approximated keeping in view of this  
inequality).
However a simple calculation shows that the effect of this 
incoherent  contribution makes its presence felt in the next order. 
The quantum noise due to surroundings becomes appreciable only at very low 
temperature.

(iii) We have already pointed out that Eq.(3) because of Born approximation 
is valid for weak damping case. We take care of this limitation
by choosing small values of $\gamma$ for carrying out numerical simulation
of quantum master equation (3) and comparing our results with semiclassical
analysis. The latter analysis based on Eq.(4) is free from Born approximation
and is therefore valid for both weak and strong damping limits. For a 
comparison over the entire range of dissipation one needs simulation of 
other kinds of master equation which are free from weak coupling approximation.
Unfortunately most of them are not well suited for numerical implementation.

We thus summarise the main conclusions of this study;

(i) The fluctuation of the second derivative of the potential due to classical
chaos is amenable to a stochastic description when the correlation time 
of fluctuations is short but finite.

(ii) $\hbar$-scaling identifies an early stage of quantum evolution which is 
dominated by chaotic diffusion and dissipation but not by thermal diffusion.

(iii) The drift and diffusion terms of the Fokker-Planck equation are intrinsic
characteristic of dynamical properties of the system since they depend 
crucially on the correlation of the fluctuations of the curvature of the 
potential. The dissipation is due to the coupling of the system to the 
external reservoir which causes irreversible evolution and is truely a 
many-body effect. On the other hand the chaotic diffusion imparts a kind
of irreversibility in the evolution which has a strict deterministic 
origin and is characteristic of the nonlinear system itself.

(iv) The Fokker-Planck equation is reminiscent of Kramers' equation which 
describes the Brownian motion in phase space for thermally activated 
processes. The Fokker-Planck equation also assumes a generic form for 
two-degree-of-freedom systems, in general.

(v) Our results show how the initial quantum noise gets amplified by classical
chaotic diffusion and then ultimately equilibriated with the passage of time
under the influence of dissipation.

(vi) We establish that there exists a critical limit to the expansion of the 
phase space which is determined by chaotic diffusion and 
dissipation.

Henon-Heiles system is a classic Hamiltonian that illustrates deterministic
stochasticity
in two-degree-of-freedom systems. In view of its prototypical role played in 
earlier as well as in the present investigation, we hope that the conclusions
drawn here will find qualititative and semiquantative applicability in other
cases of dissipative two-degree-of freedom systems at the semiclassical
level of description, in general.

\vspace{0.5cm}

\acknowledgments
B. C. Bag is indebted to the Council of Scientific and
Industrial Research for a fellowship. Partial financial support from the
Department of Science and Technology, Govt. of India, is thankfully acknowledged.  

\newpage

\begin{appendix}

\section{The transformation of the Fokker-Planck equation}
  
The diffusion terms corresponding to (38) under weak noise-approximation are given by 

\begin{eqnarray}
E_1' & = & A_4\eta_2^2(0)+C_4\eta_1^2(0)-\eta_1(0)\eta_2(0)(B_4'+B_4) 
\; \; ,\nonumber \\
E_2' & = & A_4\eta_1^2(0)+\eta_1(0)\eta_2(0)(B_4+B_4')+C_4\eta_2^2(0)
\; \; ,\nonumber \\
F_1' & = & \eta_1(0)\eta_2(0)(A_4-C_4)+B_4\eta_2^2(0)-B_4'\eta_1^2(0) 
\; \; ,\nonumber \\
F_2' & = & \eta_1(0)\eta_2(0)(A_4-C_4)-B_4\eta_1^2(0)+B_4'\eta_2^2(0)
\; \; ,\nonumber \\
G' & = & \eta_2^2(0)(A_2+B_2)+\eta_1^2(0)(C_2-B_2')
+\eta_1(0)\eta_2(0)(A_2-C_2
-B_2'-B_2) \nonumber \\
& & + \eta_1(0)\eta_4(0)(B_3'+C_3)+\eta_2(0)\eta_3(0)(B_3-A_3)
+\eta_1(0)\eta_3(0)(B_3'-C_3) \nonumber \\
& & -\eta_2(0)\eta_4(0)(A_3+B_3) \; \; , \nonumber \\
H' & = & \eta_2^2(0)(B_2'+C_2)+\eta_1^2(0)(A_2-B_2)
+\eta_1(0)\eta_2(0)(A_2-C_2
+B_2+B_2')  \nonumber \\
& & - \eta_1(0)\eta_4(0)(A_3+B_3)+\eta_1(0)\eta_3(0)(B_3-A_3)-\eta_2(0)\eta_4(0)(B_3'-C_3)
\nonumber \\
& & + \eta_2(0)\eta_3(0)(C_3-B_3') \; \;.
\end{eqnarray}

Zeroes in  $\eta_1(0), \eta_2(0), \eta_3(0)$ and $\eta_4(0)$    
refer to their initial values corresponding to the 
initial preparation of the coherent wave packet which is centered around the 
classical position and momentum for the chaotic trajectory.

The Fokker-Planck 
equation (37) can then be written in a more compact form as follows;
\begin{eqnarray}
\frac{\partial \langle \phi \rangle}{\partial t} & = & [-\eta_3
\frac{\partial \langle \phi \rangle}{\partial \eta_1}-\eta_4\frac{\partial
\langle \phi \rangle}{\partial \eta_2}+(\eta_1k+\eta_2l+2\gamma \eta_3)
\frac{\partial \langle \phi \rangle}{\partial \eta_3}  \nonumber \\
& + & (\eta_1l+\eta_2 m+ 
2\gamma \eta_4)\frac{\partial}{\partial \eta_4} \langle \phi \rangle
+4\gamma \langle \phi \rangle
+E_1' \frac{\partial^2 \langle \phi \rangle}{\partial \eta_3 
\partial \eta_1}+ E_2'\frac{\partial^2 \langle \phi \rangle}{\partial \eta_4 
\partial \eta_2} \nonumber \\
& + & F_1'\frac{\partial^2 \langle \phi \rangle}
{\partial \eta_3 \partial \eta_2}+F_2'\frac{\partial^2 \langle \phi \rangle}  
{\partial \eta_4 \partial \eta_1}+
G'(\frac{\partial^2 }{\partial \eta_3^2}+\frac{\partial^2 }
{\partial \eta_3 \partial \eta_4 }) \langle \phi \rangle \nonumber \\
& + & H'(\frac{\partial^2}{\partial \eta_4 \partial \eta_3}
+\frac{\partial^2}{\partial \eta_4^2}) ] \langle \phi \rangle   \; \; \; .
\end{eqnarray}
where
\begin{eqnarray}
k & = &1-C_3+B_3'-B_3-A_3+\langle \zeta_1 \rangle \; \; \;, \nonumber\\
m & = &1-\langle \zeta_1 \rangle+B_3-B_3'-A_3-C_3 \; \; \;, \nonumber\\
l & = &-B_3'-A_3-C_3+B_3-\langle \zeta_2 \rangle \; \;.
\end{eqnarray}

We now make the following
transformation;

\begin{equation}
u = a \eta_1+b \eta_2 +c \eta_3 +\eta_4 
\end{equation}

where a, b and c are constants to be determined.
Using this transformation we can write the Eq(A2) as

\begin{equation}
\frac{\partial \langle \phi \rangle}{\partial t}
 =  [ \lambda u \frac{\partial}{\partial u}+ A \frac{\partial^2}{\partial u^2} 
 +4 \gamma ] \langle \phi \rangle
\end{equation}

where

\begin{equation}
A = E_1'ac+E_2'b+F_1'bc+F_2'a+G'(c^2+c)+H'(c+1) 
\end{equation}

and

\begin{equation}
\lambda u =  -\eta_3a-\eta_4b+\eta_1ck+\eta_2cl 
 +2\gamma \eta_3 c+\eta_1l+\eta_2 m+2\gamma \eta_4 \; \; \;. 
\end{equation}

Making use of equation (A4) in (A7) we obtain, 

\begin{eqnarray}
\lambda a =ck+l \; \; , \nonumber\\
\lambda b =cl+m \; \; , \nonumber\\
\lambda c =-a+2\gamma c \; \; , \nonumber\\
\lambda =2\gamma-b \; \; .
\end{eqnarray}

The above relations can be used to obtain the following algebraic equation for
$\lambda$
\begin{equation}
\lambda^4+d_1 \lambda^3+d_2 \lambda^2+d_3 \lambda+d =  0  \; \; \;, \nonumber
\end{equation}

where
\begin{eqnarray}
d_1 &= &-4\gamma \; \; \;, \nonumber\\
d_2 &= & 4\gamma^2+k+m \; \; \;, \nonumber\\
d_3 & = & -2\gamma(k+m)   \; \; \;, \nonumber\\
d &= &m k-l^2 \; \; \;.
\end{eqnarray}

We now seek for a perturbative solution of Eq.(A9). To this end let us first
note that in the limit $\gamma \rightarrow 0$ Eq.(A9) reduces to a biqudratic
form (since $d_1$ and $d_3$ vanishes) whose solution is given by

\begin{equation}             
\lambda_0 = \pm
\left[\frac{-d_2 ' \pm \sqrt{{d_2'}^2-4d}}{2}
\right]^{\frac{1}{2}}\; \; \nonumber\\  
\end{equation}

where
\begin{equation}
d_2' = k+m \; \;.
\end{equation}

The lowest order perturbative solution of the the Eq. (A9) is therefore given by

\begin{equation} 
\lambda =  \lambda_0
\left[1-\frac{d_1 {\lambda_0}^2+d_2+4\gamma^2 \lambda_0}
{4 {\lambda_0}^4+3d_1 {\lambda_0}^2+2d_2 \lambda_0+d_3}
\right] \; \; \;.
\end{equation}

For the present problem the positive real root of $\lambda$ is allowed which satisfies
the physical condition (the probability distribution function must vanish at
$\pm \infty$).
Now the values of the constants $a, b$ and $c$ which are used in Eq.(A4) 
can be calculated in terms of $\lambda$ which is given by Eq.(A13) as follows
\begin{eqnarray}
b & = & 2\gamma-\lambda \; \; , \nonumber\\
c &= & \frac{l}{\lambda(2\gamma-\lambda)-k} \; \; \;, \nonumber\\ 
a & = & (2\gamma-\lambda)c \; \; \;.
\end{eqnarray}

\end{appendix}

\newpage
\begin{center}
{\bf REFERENCES}
\end{center}

\vspace{0.5cm}

\begin{enumerate}
\item J. -P. Eckmann, Rev. Mod. Phys. \underline{53}, 643(1981).
\item  W. H. Louisell, Quantum Statistical Properties of Radiation
(Wiley, New York, 1973).
\item G. Gangopadhay and D. S. Ray, in Advances in Multiphoton Processes and
Spectroscopy Vol 8, edited  by S. H. Lin, A. A. Villayes and F. Fujimura, (World 
Scientific, 1993).
\item P. H\"anggi, P. Talkner and M. Borkovec, Rev. Mod. Phys. 
\underline{62}, 251 (1990).
\item  A. O. Caldeira and  A. J. Leggett, Physica \underline{121A}, 587 (1983).
\item G. W. Ford and R. F. O'Connel, Phys. Rev. Letts. \underline{77}, 798 (1996).
\item T. Dittrich and R. Graham, Phys. Rev. \underline{A42}, 4647 (1990) .
\item B. Sundaram and P. W. Milonni, Phys. Rev. \underline{E51}, 1971 (1995).
\item R. Blumel and U. Smilansky, Phys. Rev. Letts. \underline{69}, 217 (1992).
\item N. Ben-Tal, N. Moiseyev and H. J. Korsch, Phys. Rev. \underline{A46},
1669 (1992).
\item S. Chaudhuri, D. Majumdar and D. S. Ray, Phys. Rev. \underline{E53},
5816 (1996).
\item A. Su$a^\prime$rez, R. Silbey and I. Oppenheim, J. Chem. Phys. \underline{97}, 
5101 (1992).
\item A. Tameshtit and J. E. Sipe, Phys. Rev. \underline{E51}, 1582 (1995).
\item T. Shimizu, Phys. Letts. \underline{A140}, 343 (1989).
\item L. L. Bonilla and F. Guinea, Phys. Rev. \underline{A45}, 7718 (1992).
\item T. P. Spiller and J. E. Ralph, Phys. Letts. \underline{A194}, 235 (1994).
\item A. M. Kowalski, A. Plastino and A. N. Proto, Phys. Rev. \underline{E52},
165 (1995).
\item W. H. Zurek and J. P. Paz, Phys. Rev. Letts. \underline{72}, 2508 (1994).
\item R. F. Fox and T. C. Elston, Phys. Rev. \underline{E49}, 3683 (1994). 
\item D. Cohen, Phys. Rev. Letts. \underline{78}, 2878 (1997) ; Phys. Rev. 
\underline{E55}, 1422 (1997) ; J. Phys. A Math. and Gen., \underline{31}, 8199
(1998).
\item M. Henon and C. Heiles, Astro. J. \underline{69}, 73 (1964).
\item M. Toda, Phys. Letts. \underline{A48}, 335 (1974).
\item P. Brumer and J. W. Duff, J. Chem. Phys. \underline{65}, 3566 (1976).
\item S. Chaudhuri, G. Gangopadhyay and D. S. Ray, Phys. Rev. \underline{E47}, 311
(1993).
\item S. Chaudhuri, G. Gangopadhyay and D. S. Ray, Phys. Rev. \underline{E52}, 2262
(1995).
\item N. G. van Kampen, Phys. Rep. \underline{24}, 171 (1976).
\item N. G. van Kampen, Stochastic Processes in Physics, Chemistry and 
Biology, (North Holland, 1983).
\item E. P. Wigner, Phys. Rev. \underline{40}, 749 (1932).
\item G. Gangopadhyay and D. S. Ray, J. Chem. Phys. \underline{96}, 4693 (1992);
F. Haake, H. Risken, C. Savage and D. F. Walls, Phys. Rev. \underline{A34}, 3969
(1986).
\item G. Casati, B. V. Chirikov, F. M. Izrailev and J. Ford, in Stochastic
Behaviour in Classical and Quantum Hamiltonian Systems, Lecture Notes in 
Physics, Vol 93, edited by G. Casati and J. Ford (Springer-Verlag, New York, 
1979) p 334.
\item See, for example, A. J. Lichtenberg and M. A. Lieberman, Regular and
Stochastic motion (Springer-Verlag, New York, 1983).
\end{enumerate} 

\newpage

\begin{center}
{\bf Figure Captions}
\end{center}

\begin{enumerate}
\item $(a)$ Plot of $q_1$ vs $p_1$ on the Poincare suface of section for the 
Henon-Heiles sytem with damping constant $\gamma=0.001$ and initial energy 
$E=\frac{1}{8}$. $(b)$ Same as in $(a)$ but for $\gamma = 0.0$.
\item $(a)$ Same as in Fig. 1 $(a)$ but for $E=\frac{1}{6}$. $(b)$ Same as in
$(a)$  but for $\gamma = 0.0$.
\item Plot of uncertainty  product ($\Delta \eta_1 \Delta \eta_3$) 
with time for the system as in Fig.(1). The continuous line represents
the numerical simulation of the master equation (fully quantum). The dotted line refers to
semiclassical calculation ( Eq.(60)). ( Both units are arbitrary).
\item Same as in Fig.(3) but for  $E=\frac{1}{6}$.
\item Plot of evolution of entropy with time for 
damping constant $\gamma= 0.001$. (a) $E=\frac{1}{6}$ and (b) $E=\frac{1}{8}$.
\end{enumerate}

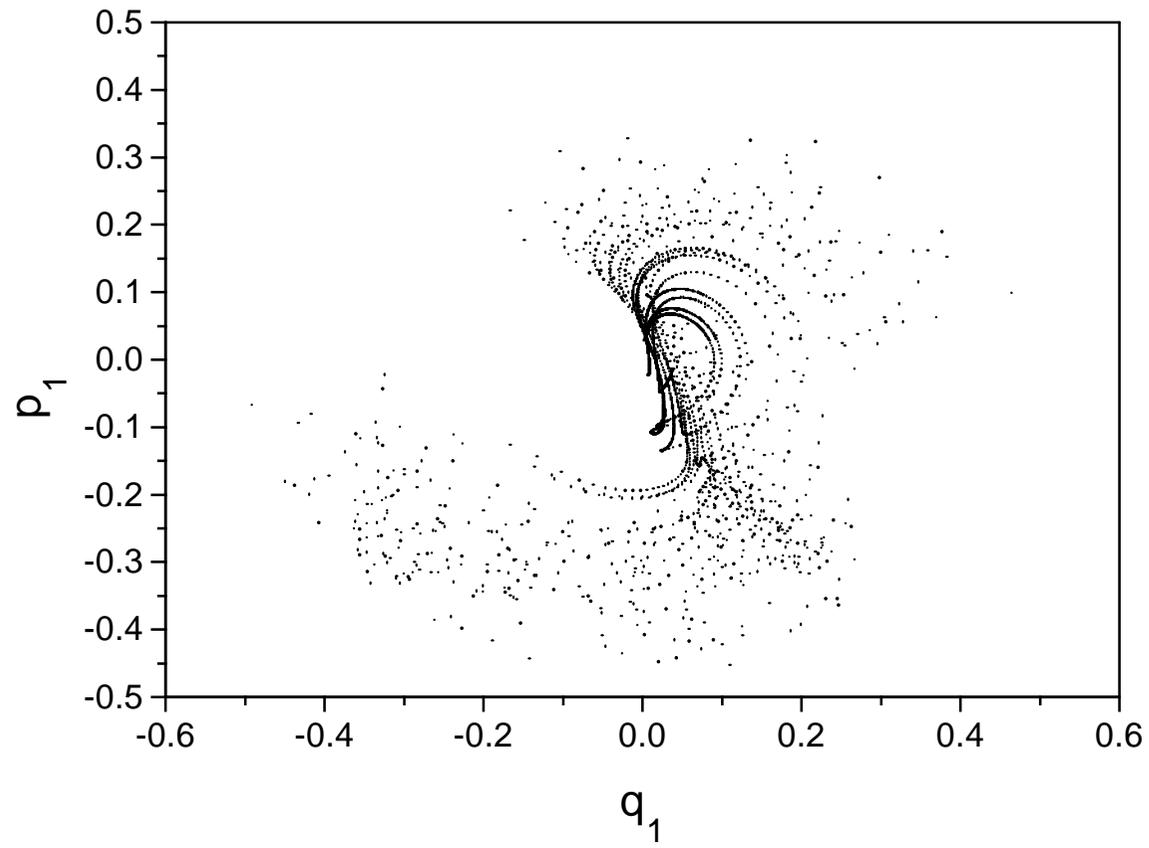
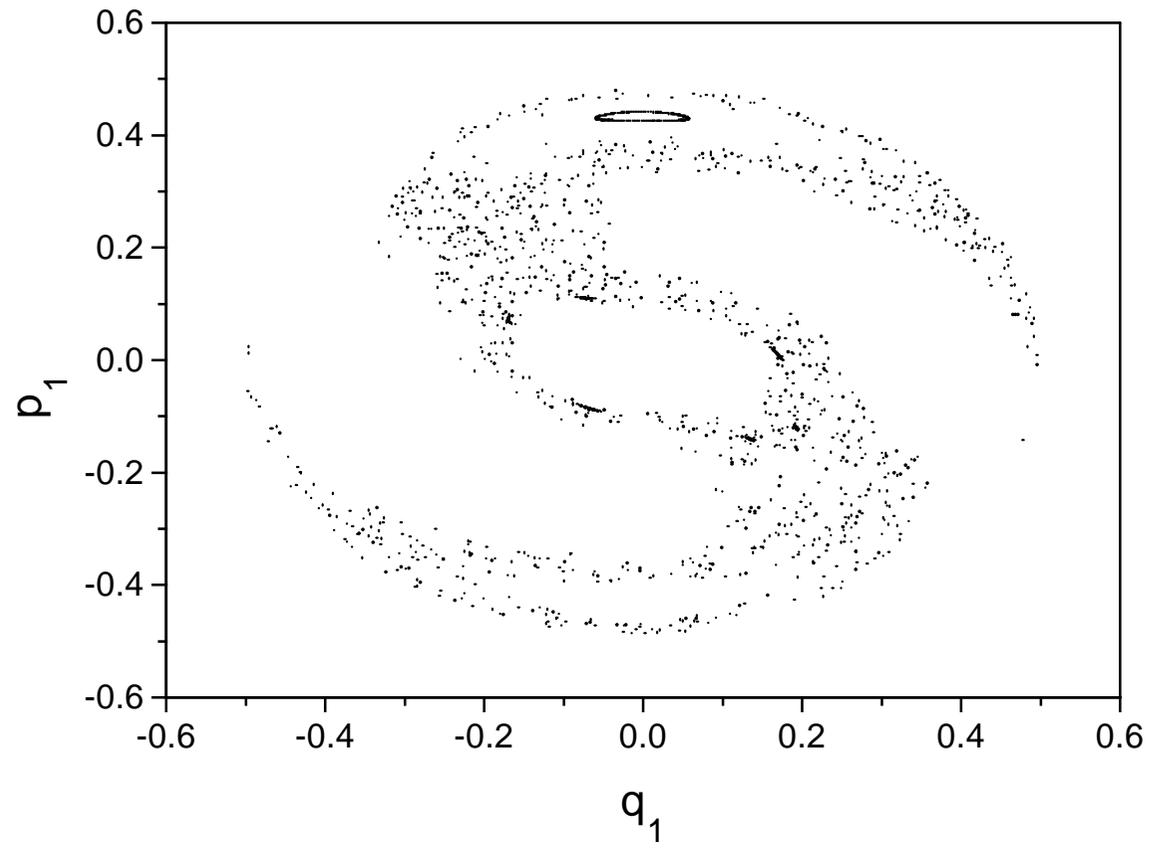
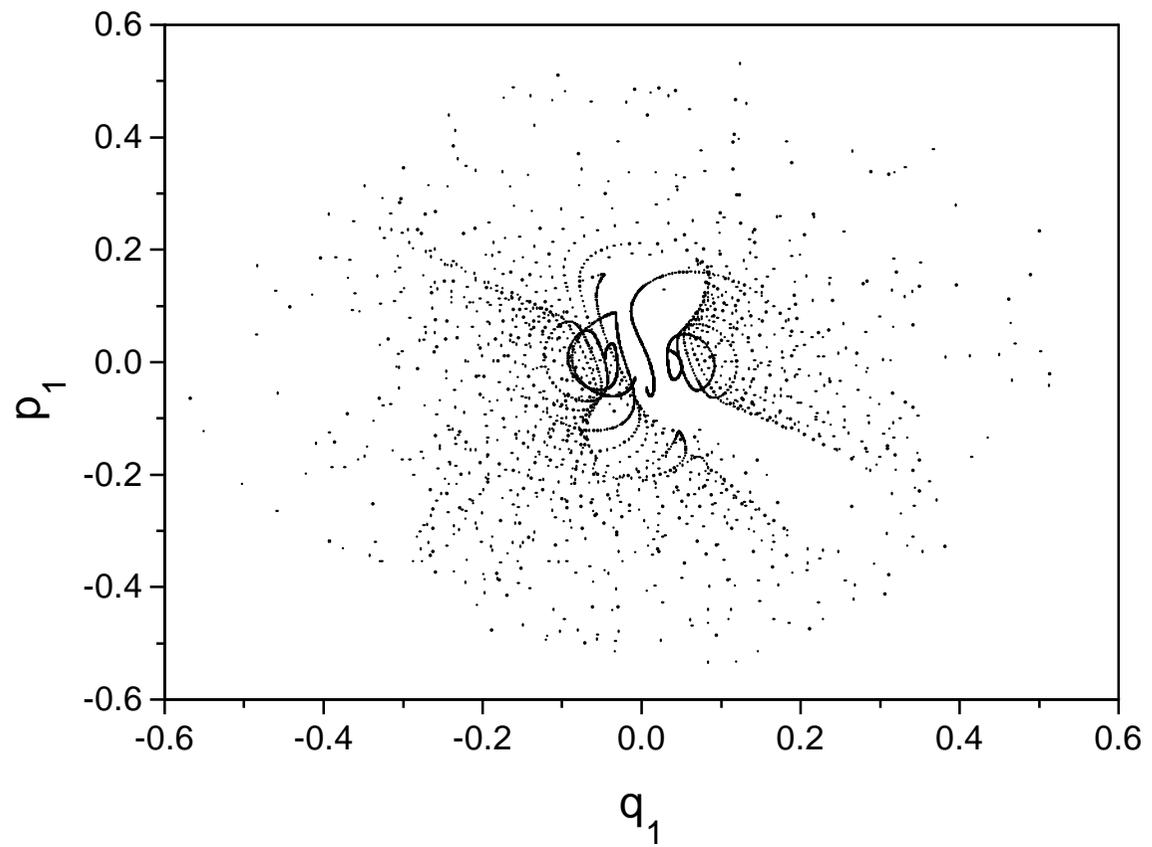
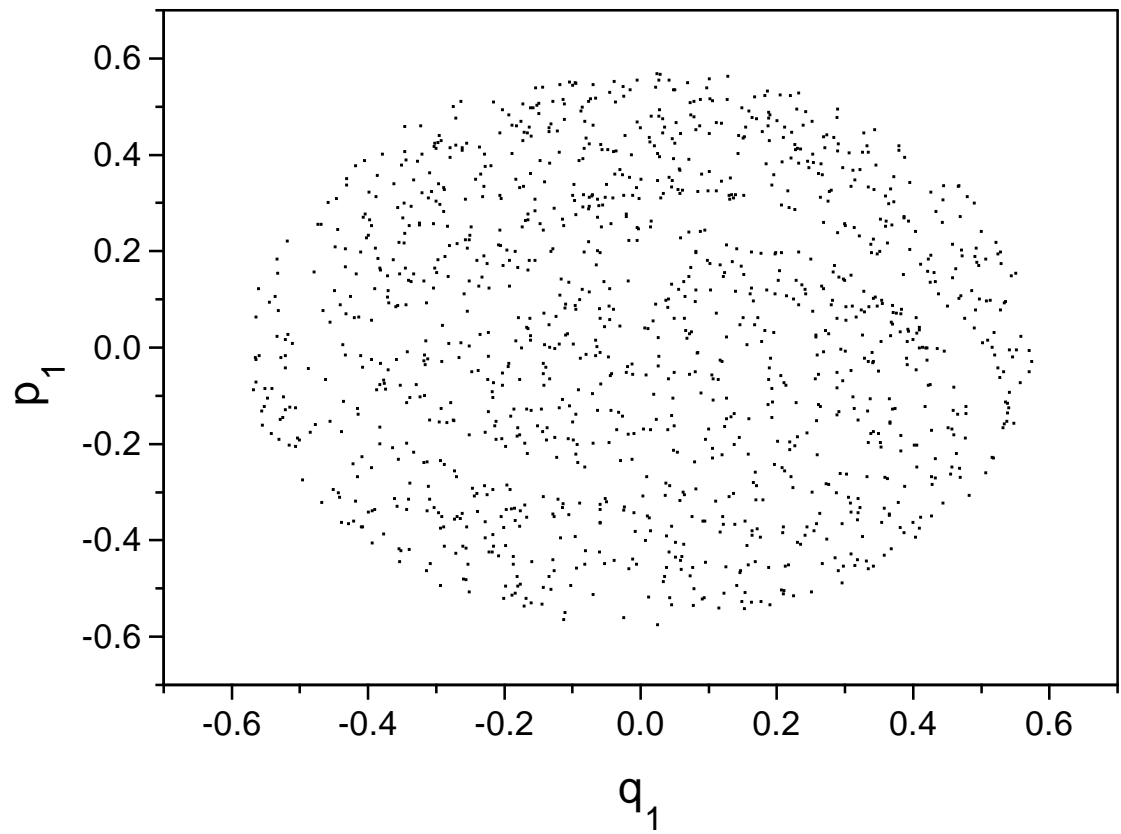
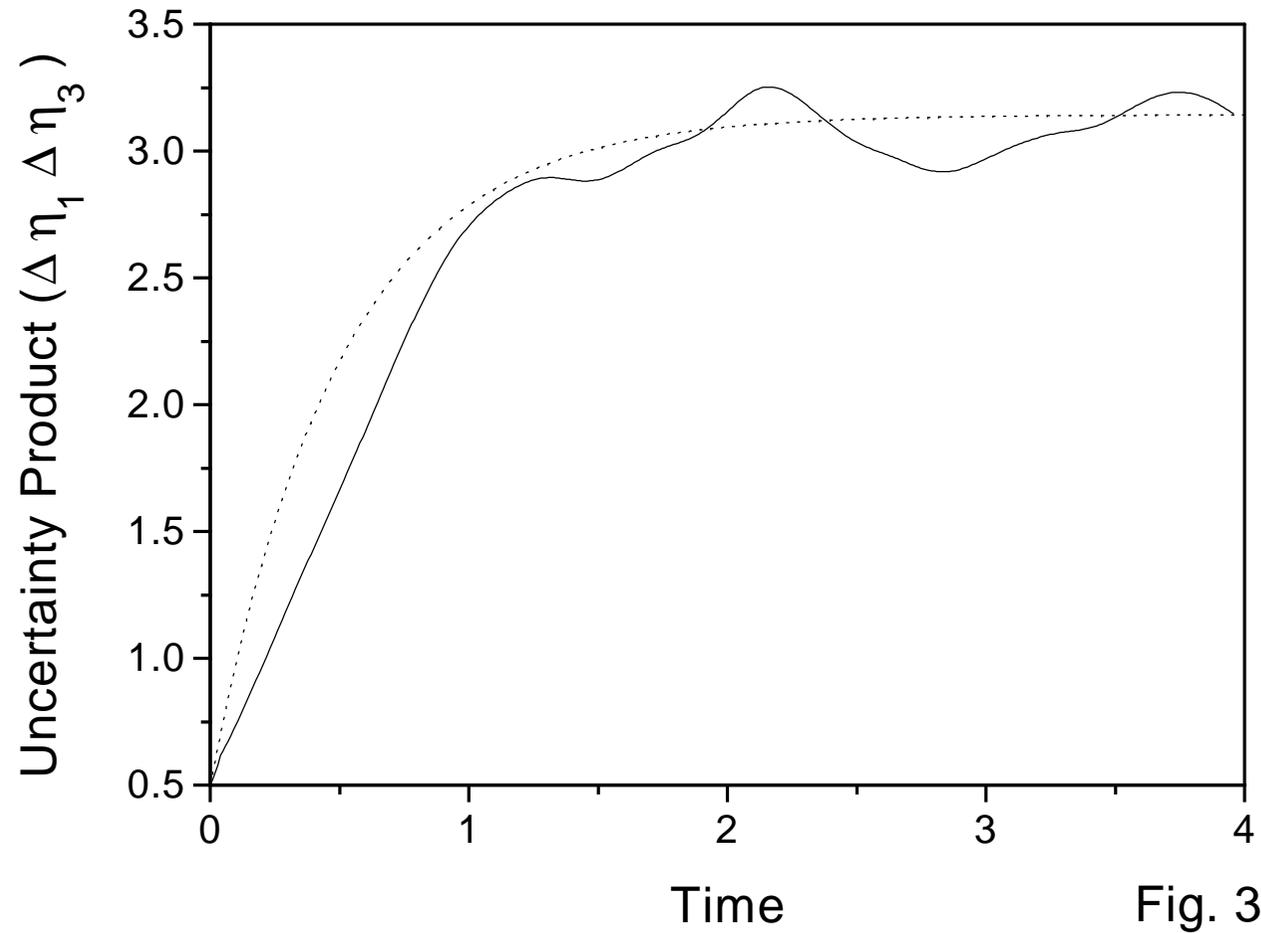
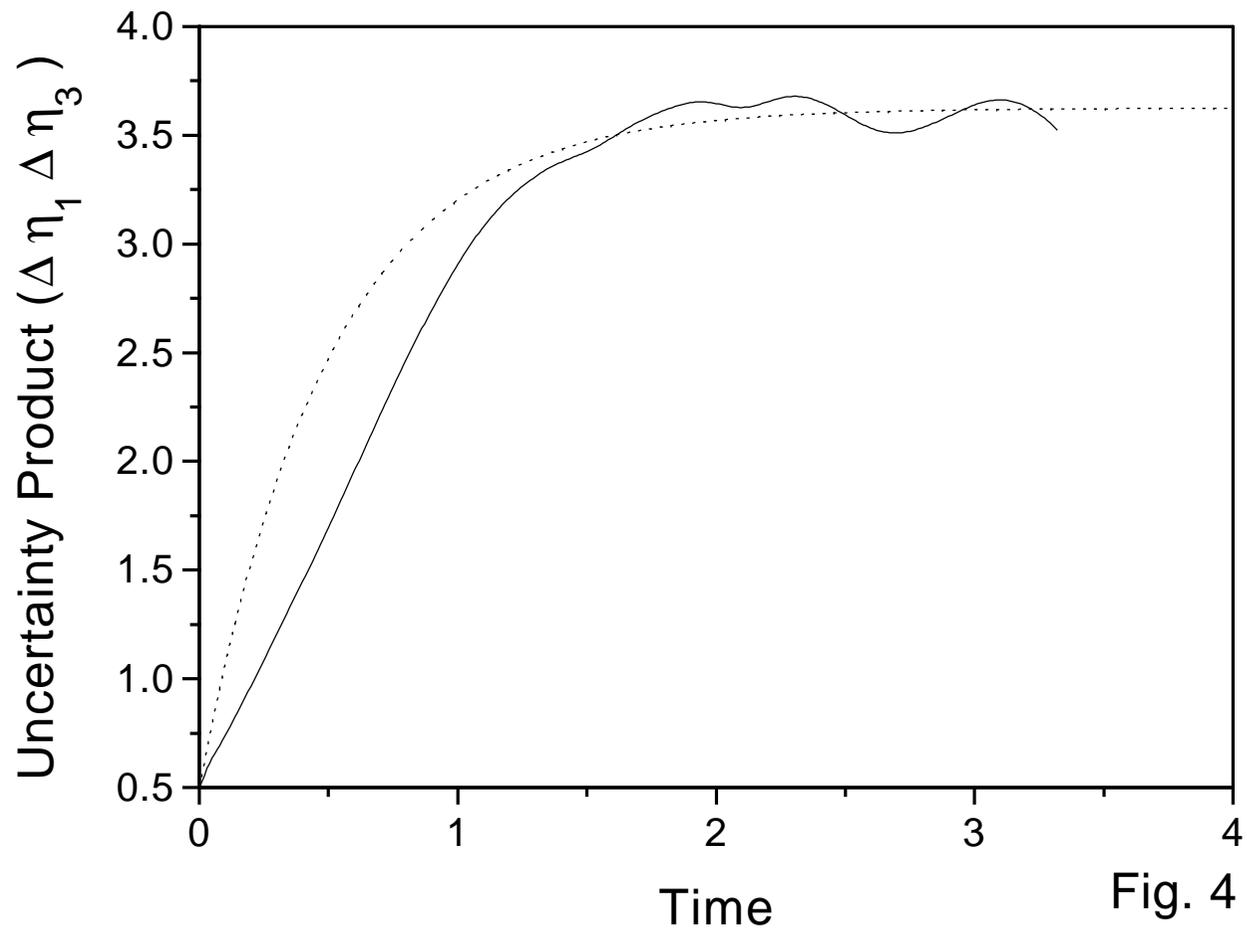
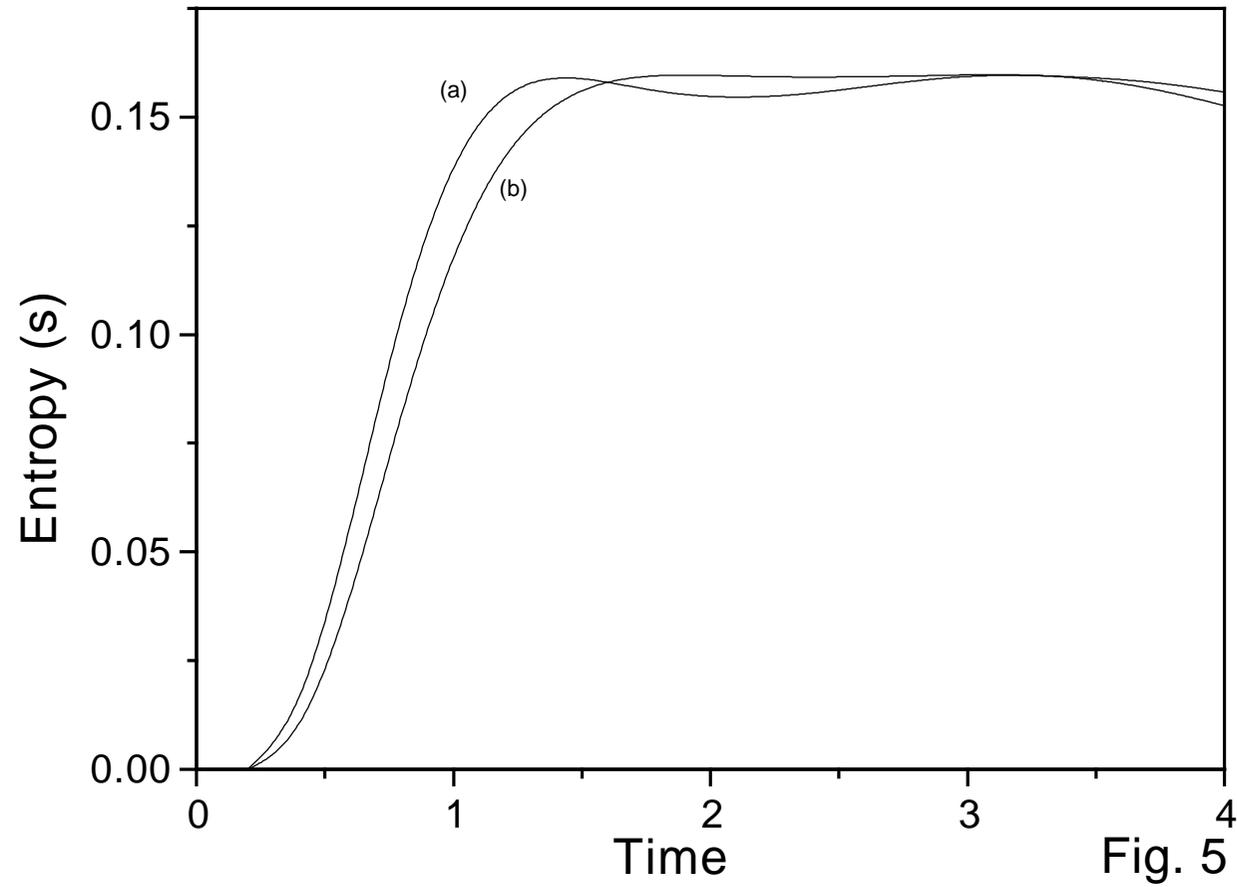
\end{document}